\documentclass[cameraready]{Interspeech}
\title{Leveraging Gradient Reversal Loss and Multitask Learning for Datasets-Aware Audio Deepfake Detection}

\author[affiliation={1}]{Mingrui}{Liang}
\author[affiliation={1}]{Thomas}{Thebaud}
\author[affiliation={2}]{\L{}ukasz}{W\'ojciak}
\author[affiliation={1}]{Laureano}{Moro Velazquez}
\author[affiliation={2}]{Yishay}{Carmiel}
\author[affiliation={1}]{Jesus}{Villalba Lopez}
\author[affiliation={1,2}]{Najim}{Dehak}
\address{
    $^1$ Department of Electrical and Computer Engineering, Johns Hopkins University, USA \\
    $^2$ Meaning, USA
}

\email{
mliang17@jh.edu,
tthebau1@jhu.edu,
}


\usepackage{comment}
\usepackage{cite}
\usepackage{amsmath,amssymb,amsfonts}
\usepackage{algorithmic}
\usepackage{graphicx}
\usepackage{textcomp}
\usepackage{xcolor}
\usepackage{adjustbox}
\usepackage{booktabs}
\usepackage{tabularx}
\usepackage{subcaption}
\usepackage{url}

\begin{document}

\maketitle

\begin{abstract}
Recent advances in speech synthesis and voice conversion, which pose threats to security and privacy, have underscored the need for deepfake detection technology. Although existing detection systems achieve strong performance on individual datasets, they often fail to generalize across diverse datasets. Prior methods for improving generalization, including data augmentation, adversarial training on auxiliary factors such as language or codec types, and Mixture-of-Experts (MoE), are limited by predefined augmentation coverage, difficulties in obtaining auxiliary factors, and substantial model complexity. In this work, we propose a practical dataset-aware framework for deepfake detection. 
% Our method focuses on adapting to unlabeled in-the-wild datasets, where auxiliary annotations such as language, codec, or spoofing method cannot be assumed to be consistently available. 
Our method targets heterogeneous datasets for which auxiliary annotations such as language, codec, or spoofing method may not be consistently available. We therefore rely only on dataset identity as a naturally available supervisory signal for multitask (MT) and gradient reversal layer (GRL) training, allowing the model to investigate both dataset-aware multitask supervision and adversarial suppression of dataset-specific information.
We conduct experiments following the 2025 Speech DeepFake Arena benchmark protocol, evaluating our model across multiple evaluation datasets and reporting aggregate performance in terms of Equal Error Rate (EER), including Average EER and Pooled EER. Compared with the baseline, MT reduces Average EER by 13.14\% relatively, while GRL reduces Pooled EER by 5.32\% relatively. These results demonstrate that our method can improve aggregate detection performance across heterogeneous evaluation datasets, offering a practical solution for deploying reliable deepfake detection systems on diverse and unseen real-world data.

% Recent advances in speech synthesis and voice conversion have increased the need for robust audio deepfake detection. Although existing systems perform well on individual datasets, they often fail to generalize across diverse and unseen data. Prior generalization methods, such as data augmentation, adversarial training with language or codec labels, and Mixture-of-Experts, are limited by predefined transformations, unavailable auxiliary annotations, or high model complexity. In this work, we propose a practical dataset-adaptive framework that uses dataset identity as a naturally available supervisory signal. We apply this signal to multitask learning (MT) and gradient reversal layer (GRL) training, enabling the model to learn dataset-related representations without requiring consistent metadata across datasets. Experiments following the 2025 Speech DeepFake Arena benchmark show that our method improves Average EER by 13.32\% and Pooled EER by 5.32\% over the baseline, demonstrating improved robustness under heterogeneous evaluation conditions.
\end{abstract}

\noindent\textbf{Project links:}

% \href{https://huggingface.co/RuiRuihigh/hyperion-mt-deepfake-detector}
{MT model repository \footnote{\url{https://huggingface.co/RuiRuihigh/hyperion-mt-deepfake-detector}}}
and
% \href{https://huggingface.co/RuiRuihigh/hyperion-grl-deepfake-detector}
{GRL model repository\footnote{\url{https://huggingface.co/RuiRuihigh/hyperion-grl-deepfake-detector}}}.

\section{Introduction}

\begin{figure}[t]
    \centering
    \includegraphics[width=\columnwidth]{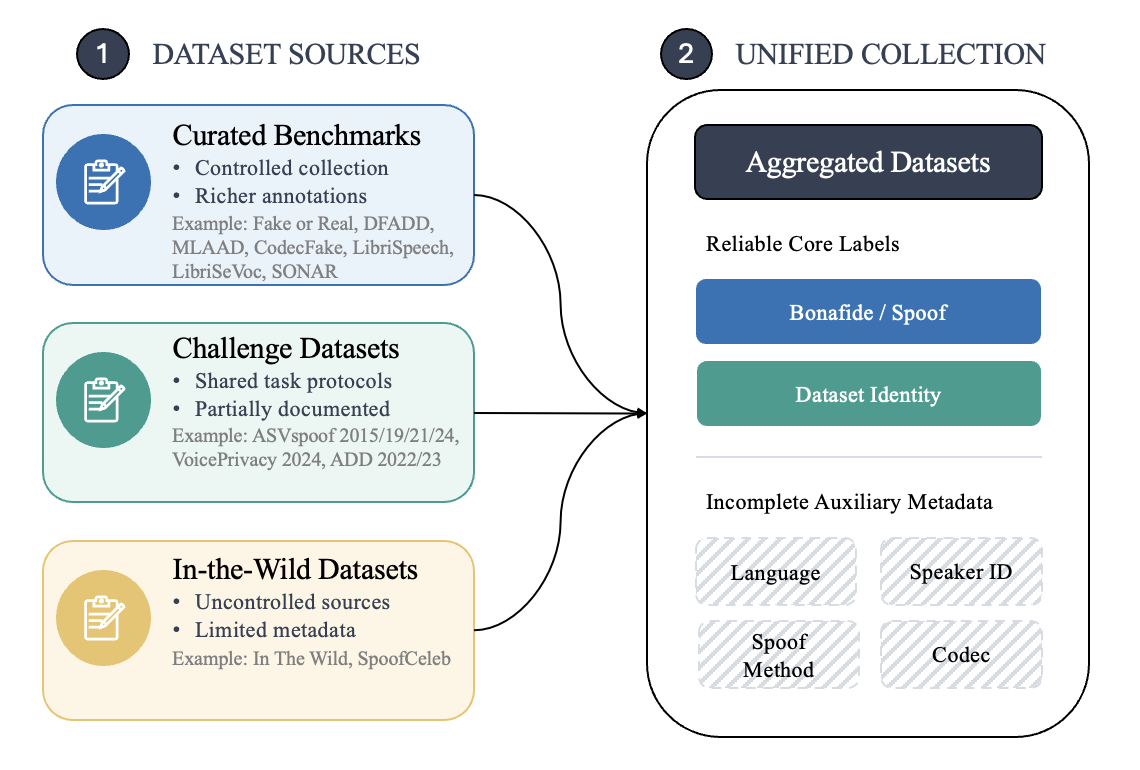}
    \caption{Overview of the unified data construction. After aggregating datasets from diverse sources, only bona fide/spoof labels and dataset identity remain consistently available across datasets. The Curated, Challenge, and In-the-Wild categories indicate the primary grouping of each dataset and are not intended to be strictly mutually exclusive.}
    \label{fig:dataset_collection}
\end{figure}

With the help of powerful deep neural networks, recent advances in text-to-speech (TTS) and voice conversion (VC) have significantly improved the perceptual quality of generated speech. Although this technique is widely used across domains, such as data enhancement \cite{yuen2023asr}, it can also be used for criminal activities, including financial fraud, political conflict, and impersonation. Therefore, audio deepfake detection has gradually become an active research field \cite{wu2023defender, khan2022voice}. Initially, researchers primarily used handcrafted features, such as CQT, STFT, and LF, as inputs to classify audio \cite{pham2025comprehensive}, achieving excellent results. Later, as the self-supervised learning (SSL) models become more and more powerful, people begin to take advantage of them in different tasks in the speech area, such as automatic speech recognition (ASR) \cite{zhao2022improving}, speaker verification (SV) \cite{jung2024espnet}, and surely audio deepfake detection (ADD) \cite{li2025survey}. By leveraging large-scale pretraining corpora and learning rich representations, SSL models significantly enhance the robustness and generalization of anti-spoofing systems \cite {chen2024singing,zhu2024slim,tak2022automatic}.
Currently, state-of-the-art systems typically combine SSL models with advanced neural backbones \cite{li2025survey}.

Although these systems have achieved remarkable performance in controlled scenarios \cite{yang2024robust, li2023voice}, the challenge of effectively generalizing their capabilities to unseen data and increasingly sophisticated speech deepfakes remains critical. This issue is further complicated by the unique characteristics of the state-of-the-art speech deepfake datasets, including diverse generation methods, multilingual data, and varied post-processing, which make them reasonable to consider as distinct domains.

Currently, the most common approach to address the domain gap is data augmentation, in which variable masks or codec types are applied to audio. Another approach is adversarial training with labels such as language or codec types, which can help the model learn language-invariant or codec-invariant information \cite{chen2021ur,ba2023transferring, zhang2021empirical}. Furthermore, MoE provides a natural way to handle domain shift by routing different experts to process data from their respective domains, thereby enabling domain-specific specialization. However, these approaches remain difficult to scale to heterogeneous training data. Data augmentation is restricted to predefined transformations and may not cover real-world dataset shifts. Adversarial training based on language or codec labels requires auxiliary annotations that are often unavailable or inconsistently defined across datasets, while MoE-based methods introduce additional model complexity and computational cost. This annotation issue becomes more evident when datasets from different sources are combined, as illustrated in Fig.~\ref{fig:dataset_collection}. In such a unified collection, metadata such as language, spoofing method, and codec type cannot be consistently obtained for all samples. In contrast, bona fide/spoof labels and dataset identity are reliably available across datasets, making dataset identity a practical supervisory signal for mixed-dataset training.

% Instead of relying on auxiliary annotations that may be unavailable or inconsistent across datasets, we use dataset identity as a naturally available supervisory signal.

Motivated by this observation, we propose a dataset-aware framework for training with heterogeneous and unlabeled in-the-wild datasets.  At the same time, our framework leverages the diversity of pooled datasets without relying on introducing substantial model complexity. Specifically, in the  multitask (MT) learning, we introduce a complex auxiliary label by combining dataset identity with the spoof/bona fide label, forming class-conditional dataset labels such as dataset-spoof and dataset-bona fide. This design ensures that the auxiliary task remains aligned with the main spoofing detection objective while still capturing dataset-specific characteristics. For the gradient reversal layer (GRL) training, we apply dataset identity directly to encourage the feature extractor to reduce dataset-specific bias. The key contributions of this work can be summarized as follows:
\begin{itemize}
    \item \textbf{Adaptation to in-the-wild datasets}: 
    Our framework uses dataset identity as a naturally available supervisory signal, avoiding the need for auxiliary annotations, which are often unavailable or inconsistent across datasets. 
    This makes newly collected or unlabeled in-the-wild datasets easier to integrate into mixed-dataset training.
    
    \item \textbf{Dataset-aware supervision}: 
    We introduce class-conditional dataset labels by combining dataset identity with bona fide/spoof labels. Compared with using dataset identity alone, this supervision better aligns the auxiliary task with the main spoofing detection objective and helps the model capture dataset-specific variations.
    
    \item \textbf{Improved detection with a lightweight single-model framework}: 
    Our approach achieves a 13.14\% relative improvement in Average EER and a 5.32\% relative improvement in Pooled EER, demonstrating stronger cross-dataset generalization. 
    Unlike MoE- or ensemble-based methods, our framework trains a single detector with 315.4M parameters, reducing training and inference complexity while maintaining strong performance.
\end{itemize}

\begin{figure*}[t]
    \centering
    \includegraphics[width=\textwidth]{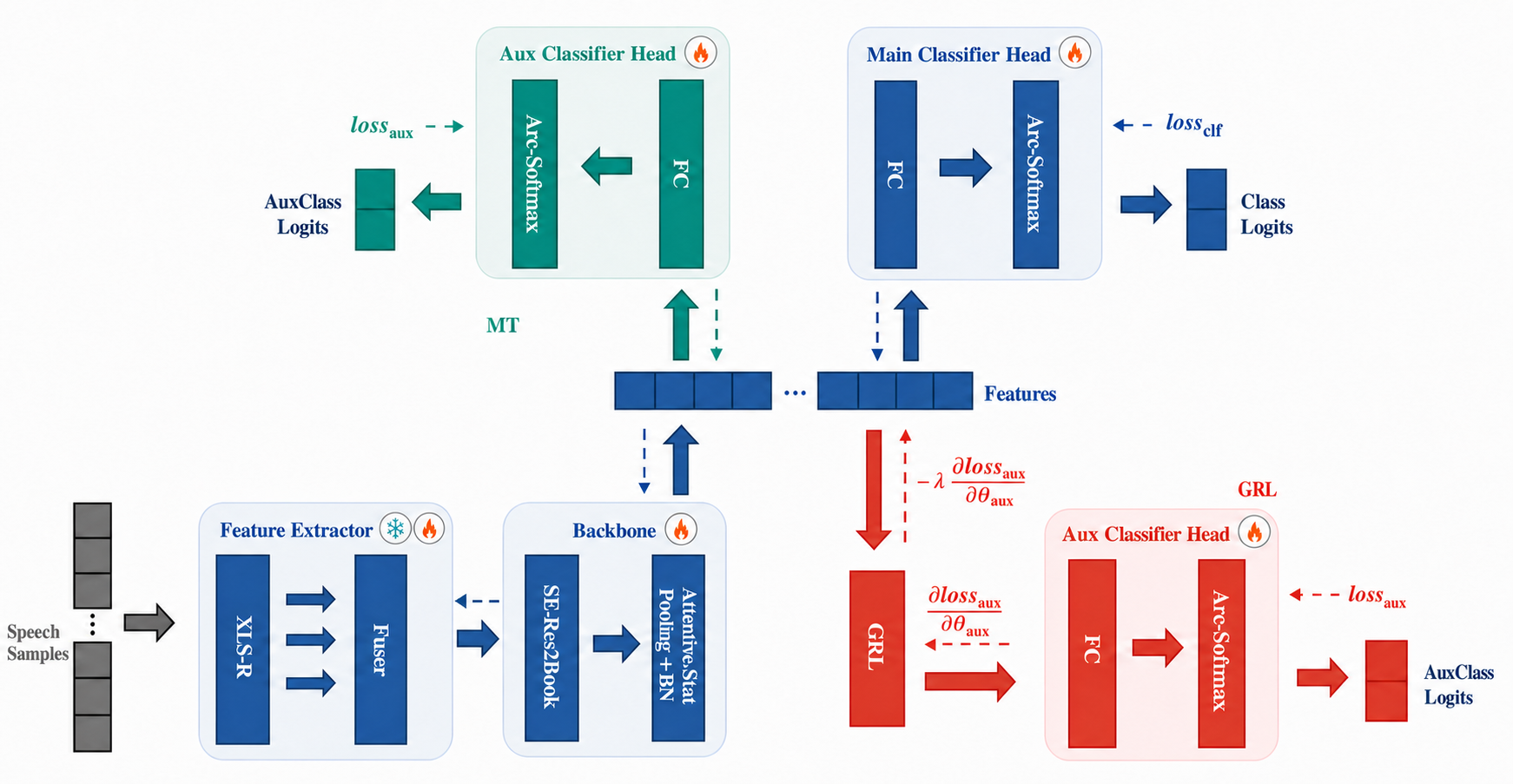}
    \caption{Overview of the proposed framework. The blue modules represent the baseline detection pipeline. MT consists of the blue modules and the green auxiliary branch for multitask learning, while GRL consists of the blue modules and the orange auxiliary branch. The snowflake and flame icons denote frozen and trainable modules, respectively, with the feature extractor frozen in Stage~1 and all modules trainable in Stage~2.}
    \label{fig:MT_GRL}
\end{figure*}

\section{Related Work}

\subsection{Hand-crafted Feature-based Audio Deepfake Detection} 
Hand-crafted features (e.g., Mel Spectrogram) are widely used across many speech-related fields. In preliminary audio deepfake detection work, these features indeed achieved some remarkable results \cite{deng2024vfd,zhang2023syntheticspeechdetectionbased}. For example, HM-Conformer \cite{shin2024hm} introduced the hierarchical pooling and multi-level classification token aggregation into Conformer \cite{gulati2020conformer}, transferring LFCC to two-class prediction. Additionally, some work used a combination of a special log-Mel spectrogram and ResNet \cite{gao2021generalized} to address deepfake detection tasks, achieving robust results.

\subsection{Self-Supervised Learning (SSL) Model-based Audio Deepfake Detection}

SSL models are usually pretrained on large amounts of unlabeled speech data with self-supervised learning techniques. Among these SSL models, the most popular models are WavLM \cite{chen2022wavlm}, HuBERT \cite{hsu2021hubert}, Wav2Vec2 2.0 \cite{baevski2020wav2vec}, and XLS-R \cite{babu2021xls}. Compared with hand-crafted features, SSL models can capture richer and more meaningful speech representations that can be effectively used in downstream tasks.

Beyond simply obtaining embeddings from the last layers of SSL, recent work leverages information from multiple layers in SSL models using various techniques \cite{kulkarni2026compact, zhang2024audio, guo2024audio, el2025comprehensive}. For instance, Zhang et al. \cite{zhang2024audio} leverage Sensitive Layer Select (SLS) to fuse representative layers automatically, and Guo et al. \cite{guo2024audio} propose using attentive statistic pooling on every layer in WavLM to improve accuracy.

With SSL models and advanced backbone models (e.g., ECAPA-TDNN \cite{desplanques2020ecapa}, BiCrossMamba-ST \cite{kheir2025bicrossmamba}, and AASIST \cite{jung2022aasist}), researchers can achieve great results. Given that XLS-R typically achieves more competitive performance \cite{ali2026superb} and that speaker information can contribute to classification \cite{zhang2023syntheticspeechdetectionbased}, we use the combination of XLS-R and ECAPA-TDNN as our baseline model.

\subsection{Domain Generalization and Robustness}

Although current models have thousands of parameters and are highly effective at discriminating real audio from fake audio, they still fall short in generalization \cite{pham2025comprehensive}. 
When these models are exposed to unseen conditions, such as novel spoofing algorithms, cross-dataset variations, and different recording channels, their performance will decline markedly. 

One way to improve robustness is to introduce data variability by data augmentation \cite{korvas_2014,snyder2015musan,tak2022rawboost}. For instance, RawBoost \cite{tak2022rawboost} enhances robustness by applying a series of realistic signal-level perturbations directly to raw waveforms, including channel effects, additive noise, and nonlinear distortions. 

Another direction focuses on learning domain-invariant representations through multitasking and adversarial domain adaptation. Specifically, multitasking explicitly retains domain-specific information, whereas adversarial domain adaptation aims to align feature distributions across domains by removing domain-specific factors. Previous work has primarily leveraged auxiliary classifiers and gradient reversal on auxiliary labels such as language, device, and codec types \cite{chen2021ur,ba2023transferring, zhang2021empirical}.

More recently, model specialization strategies have been explored to improve generalization. Negroni et al. propose a Mixture-of-Experts (MoE) framework, where multiple domain-specific detectors are combined through a gating mechanism to dynamically adapt to different input conditions \cite{negroni2025leveraging}. This approach leverages domain-specific knowledge and improves adaptability to diverse datasets.

Despite their effectiveness, these approaches remain limited in real-world scenarios. Data augmentation improves robustness by increasing data diversity, but it does not fundamentally address unseen distributions, as deepfake techniques continue to evolve \cite{li2025survey}. 
Existing MT and adversarial adaptation methods often use auxiliary metadata, such as language, codec, or synthesis method, as supervision. However, these labels are frequently unavailable or inconsistently defined across datasets, limiting their applicability in large-scale mixed-dataset training.
Meanwhile, model specialization approaches such as Mixture-of-Experts require training multiple large models and additional gating mechanisms, leading to high computational cost and poor scalability \cite{feng2025dive}. These limitations highlight the need for more flexible and efficient approaches for domain generalization.

\begin{table*}[t]
\centering
\caption{Summary of the datasets used for training, validation, and evaluation, including language coverage and the total number of utterances.}
\label{tab:dataset_usage}

\small
\setlength{\tabcolsep}{13pt}          % 增加横向列间距
\renewcommand{\arraystretch}{1.08}    % 增加纵向行距

\begin{tabular}{lccccc}
\toprule
\textbf{Dataset} & \textbf{Language} & \textbf{Total \#utt} &
\textbf{Train} & \textbf{Dev} & \textbf{Eval} \\
\midrule
ASVspoof 2015~\cite{wu2017asvspoof}          & English      & 263{,}151     & \checkmark & \checkmark & \\
ASVspoof 2019~\cite{todisco2019asvspoof}     & English      & 121{,}461     & \checkmark & \checkmark & \checkmark \\
ASVspoof 2024~\cite{wang2024asvspoof}        & English      & 1{,}004{,}078 & \checkmark & \checkmark & \checkmark \\
Fake or Real~\cite{reimao2019dataset}        & English      & 37{,}541      & \checkmark & \checkmark & \checkmark \\
DFADD~\cite{du2024dfadd}                     & English      & 191{,}039     & \checkmark & \checkmark & \checkmark \\
MLAAD~\cite{muller2024mlaad}                 & Multilingual & 54{,}000      & \checkmark & \checkmark & \\
Codecfake~\cite{xie2025codecfake}            & English      & 1{,}172{,}314 & \checkmark & \checkmark & \checkmark \\
SpoofCeleb~\cite{jung2025spoofceleb}         & English      & 2{,}631{,}551 & \checkmark & \checkmark & \\
LibriSpeech~\cite{panayotov2015librispeech}  & English      & 281{,}241     & \checkmark & \checkmark & \\
VoicePrivacy 2024~\cite{tomashenko2025first} & English      & 728{,}098     & \checkmark & \checkmark & \\
SONAR~\cite{li2024sonar}                     & Multilingual & 3{,}947       & \checkmark & \checkmark & \checkmark \\
LibriSeVoc~\cite{sun2023ai}                  & English      & 92{,}407      & \checkmark & \checkmark & \checkmark \\
ASVspoof 2021~\cite{liu2023asvspoof}         & English      & 181{,}566     &            &            & \checkmark \\
In The Wild~\cite{muller2022does}            & English      & 31{,}779      &            &            & \checkmark \\
ADD 2022~\cite{yi2022add}                    & Chinese      & 404{,}329     &            &            & \checkmark \\
ADD 2023~\cite{yi2023add}                    & Chinese      & 285{,}861     &            &            & \checkmark \\
\bottomrule
\end{tabular}

\vspace{0.35em}
\begin{minipage}{0.94\textwidth}
\footnotesize
\textit{Note:} Dev denotes the validation split, and Total \#utt denotes
the sum of all available utterances for each dataset. For ADD 2022 and
ADD 2023, all provided tasks and evaluation rounds are included.
\end{minipage}
\end{table*}

\section{Method}
To make the system robust across scenarios, the key is to enable it to learn characteristics that are irrelevant to those scenarios, such as dataset-specific ones. Considering this, we propose two methods to explicitly allow the system to understand the special information: multitask (MT) and gradient reversal layer (GRL).
% \begin{figure}[t]
%     \centering
%     \includegraphics[width=0.9\linewidth]{figures/Baseline.png}
%     \caption{Baseline Pipeline.}
%     \label{fig:Baseline}
% \end{figure}

% \begin{figure*}[t]
%     \centering
%     \includegraphics[width=\textwidth]{figures/ConditionInjection.png}
%     \caption{Condition Injection system pipeline.}
%     \label{fig:ConditionInjection}
% \end{figure*}
% \begin{figure}[t]
%     \centering
%     \includegraphics[width=0.9\linewidth]{figures/GatedTCAC.png}
%     \caption{Adapter Block (GatedTCACs).}
%     \label{fig:GatedTCAC}
% \end{figure}

Our baseline builds on the SSL-based spoofing detection framework explored in~\cite{kulkarni2024exploring}, where SSL front-end representations combined with an ECAPA-TDNN backbone are shown to be effective for generalizing to unseen audio data. As shown in the blue modules of Fig.~\ref{fig:MT_GRL}, raw audio waveforms are first processed by an XLS-R encoder to extract high-level representations. Beyond simply using the encoder's last layer, we employ attentive fusion to integrate outputs from multiple encoder layers. The fused feature is then fed into a convolutional backbone network, where we modify the ECAPA-TDNN from three SE-Res2Blocks to a single SE-Res2Block to capture both temporal and frequency-related dependencies. An attentive statistical pooling layer is applied to aggregate variable frame-level features into a fixed embedding. Finally, a fully connected classifier head is used to predict the binary label, spoof or bona fide. Totally, the whole framework only contains 315.4M parameters.

\subsection{Multitask Learning}
\label{sec:mt}

In MT (the blue and green modules) of Fig.~\ref{fig:MT_GRL}, we adopt a multitask learning strategy to explicitly model dataset-dependent variations. 
For each input sample $x$, we denote its spoofing label as $y \in \{\text{bona fide}, \text{spoof}\}$ and its dataset identity as $d$. 
Instead of using the dataset identity alone as the auxiliary label, we construct a class-conditional dataset label:
\begin{equation}
y_{\text{aux}}^{\text{MT}} = (d, y),
\end{equation}
which corresponds to labels such as dataset-bona fide and dataset-spoof.

Given an input sample $x$, the backbone extracts a feature representation:
\begin{equation}
f = \mathcal{F}(x).
\end{equation}
The main classifier predicts the binary spoofing label:
\begin{equation}
z_{\text{main}} = \mathcal{C}_{\text{main}}(f),
\end{equation}
while the auxiliary classifier predicts the class-conditional dataset label:
\begin{equation}
z_{\text{aux}}^{\text{MT}} = \mathcal{C}_{\text{aux}}^{\text{MT}}(f).
\end{equation}
The corresponding losses are:
\begin{equation}
\mathcal{L}_{\text{main}} =
\text{CE}(z_{\text{main}}, y),
\end{equation}
\begin{equation}
\mathcal{L}_{\text{aux}}^{\text{MT}} =
\text{CE}(z_{\text{aux}}^{\text{MT}}, y_{\text{aux}}^{\text{MT}}),
\end{equation}
where $\text{CE}(\cdot)$ denotes the cross-entropy loss.
The overall multitask objective is:
\begin{equation}
\mathcal{L}_{\text{MT}} =
\mathcal{L}_{\text{main}}
+
\lambda_{\text{MT}}
\mathcal{L}_{\text{aux}}^{\text{MT}},
\end{equation}
where $\lambda_{\text{MT}}$ controls the contribution of the auxiliary task.

This class-conditional auxiliary supervision encourages the representation to capture dataset-specific variations while remaining aligned with the main spoofing detection objective. 
Compared with predicting dataset identity alone, our proposed label reduces ambiguity because the auxiliary task is conditioned on the bona fide/spoof distinction rather than being independent of it.

\subsection{Gradient Reversal Layer}
\label{sec:grl}

In GRL (the blue and orange modules) of Fig.~\ref{fig:MT_GRL}, we adopt a gradient reversal layer to learn dataset-invariant representations. 
Unlike multitask learning, which encourages the backbone to encode dataset-related information, GRL discourages the backbone from preserving information that allows the dataset identity to be predicted.

Given the feature representation:
\begin{equation}
f = \mathcal{F}(x),
\end{equation}
we pass it through a GRL before feeding it into the auxiliary dataset classifier:
\begin{equation}
\tilde{f} = \text{GRL}(f),
\end{equation}
\begin{equation}
z_{\text{aux}}^{\text{GRL}} =
\mathcal{C}_{\text{aux}}^{\text{GRL}}(\tilde{f}).
\end{equation}
For GRL training, the auxiliary label is the dataset identity:
\begin{equation}
y_{\text{aux}}^{\text{GRL}} = d.
\end{equation}
The auxiliary loss is therefore:
\begin{equation}
\mathcal{L}_{\text{aux}}^{\text{GRL}} =
\text{CE}(z_{\text{aux}}^{\text{GRL}}, d).
\end{equation}
During the forward pass, the GRL acts as an identity function:
\begin{equation}
\tilde{f} = \text{GRL}(f) = f.
\end{equation}
During backpropagation, however, it reverses the gradient passed to the backbone:
\begin{equation}
\frac{\partial \mathcal{L}_{\text{aux}}^{\text{GRL}}}{\partial f}
=
-
\frac{\partial \mathcal{L}_{\text{aux}}^{\text{GRL}}}{\partial \tilde{f}}.
\end{equation}
The training loss is written as:
\begin{equation}
\mathcal{L}_{\text{GRL}} =
\mathcal{L}_{\text{main}}
+
\lambda_{\text{GRL}}
\mathcal{L}_{\text{aux}}^{\text{GRL}},
\end{equation}
where $\lambda_{\text{GRL}}$ controls the strength of the auxiliary objective. 
The adversarial effect is introduced by the GRL during backpropagation, rather than by explicitly subtracting the auxiliary loss from the training objective.

Specifically, the auxiliary classifier is optimized normally to predict the dataset identity. 
However, for the backbone, the effective gradient becomes:
\begin{equation}
\frac{\partial \mathcal{L}_{\text{GRL}}}{\partial \theta_{\mathcal{F}}}
=
\frac{\partial \mathcal{L}_{\text{main}}}{\partial \theta_{\mathcal{F}}}
-
\lambda_{\text{GRL}}
\frac{\partial \mathcal{L}_{\text{aux}}^{\text{GRL}}}{\partial \theta_{\mathcal{F}}},
\end{equation}
where $\theta_{\mathcal{F}}$ denotes the parameters of the backbone.

Therefore, the auxiliary classifier learns to identify the dataset source, while the backbone learns to make this prediction difficult. 
This encourages the learned representation to remove dataset-discriminative information, thereby promoting dataset-invariant features and improving cross-dataset generalization.

\section{Experiment}
\subsection{Datasets}

% For training and validation, we construct a large-scale unified dataset by combining multiple publicly available corpora, including ASVspoof 2015, ASVspoof 2019, ASVspoof 2024, Fake or Real, DFADD, MLAAD, Codecfake, Spoofceleb, Librispeech, VoicePrivacy 2024, SONAR and LibriSeVoc. These datasets cover a wide range of spoofing techniques, recording conditions, and languages, providing diverse supervision for robust model training.

We construct a large-scale unified corpus for training and validation by combining multiple publicly available datasets. The usage of each dataset across the training, validation, and evaluation splits is summarized in Table~\ref{tab:dataset_usage}. Specifically, for training and validation, we have ASVspoof 2015, ASVspoof 2019, ASVspoof 2024, Fake or Real, DFADD, MLAAD, Codecfake, Spoofceleb, Librispeech, VoicePrivacy 2024, SONAR and LibriSeVoc. These corpora cover diverse spoofing techniques, recording conditions, speaker characteristics, and languages, providing broad supervision for robust speech deepfake detection.

The resulting training subset contains 7,325.65 hours of audio and 4,859,430 utterances in total. Among them, 4,198,675 are spoof samples and 660,755 are bona fide samples, accounting for 86.4\% and 13.6\% of the training utterances, respectively.

% For evaluation, we follow the benchmark protocol of the 2025 Speech Deepfake Arena \cite{dowerah2026speech}, where the evaluation datasets include In The Wild \cite{muller2022does}, ASV2019, ASV2021LA \cite{liu2023asvspoof}, ASV2021DF, ASV2024-EVAL, Fake or Real, Codecfake, ADD 2022 Track1 \cite{yi2022add}, ADD 2022 Track2, ADD 2023 R1 \cite{yi2023add}, ADD 2023 R2, DFADD, LibriSeVoc, SONAR.

% For evaluation, we follow the benchmark protocol of the 2025 Speech Deepfake Arena~\cite{dowerah2026speech}, where the evaluation datasets include In The Wild, ASVspoof 2019, ASVspoof 2021LA, ASVspoof 2021DF, ASVspoof 2024, Fake or Real, Codecfake, ADD 2022 Track1, ADD 2022 Track3, ADD 2023 R1, ADD 2023 R2, DFADD, LibriSeVoc, and SONAR. The evaluation suite consists of multiple in-domain and out-of-domain test sets, enabling a comprehensive assessment of model generalization across heterogeneous speech deepfake scenarios.
For evaluation, we follow the benchmark protocol of the 2025 Speech Deepfake Arena~\cite{dowerah2026speech}, where the evaluation datasets include In The Wild, ASVspoof 2019, ASVspoof 2021, ASVspoof 2024, Fake or Real, Codecfake, ADD 2022, ADD 2023, DFADD, LibriSeVoc, and SONAR. The evaluation suite consists of multiple in-domain and out-of-domain test sets, enabling a comprehensive assessment of model generalization across heterogeneous speech deepfake scenarios.

% Based on the overlap with the training datasets, we further divide the evaluation sets into in-domain and out-of-domain subsets. The in-domain test set comprises datasets also seen during training. In contrast, the out-of-domain test set consists of datasets that are not included in training, including In-the-Wild \cite{muller2022does}, ASVspoof 2021 \cite{liu2023asvspoof}, Fake-or-Real \cite{reimao2019dataset}, CodecFake \cite{wu2024codecfake}, DFADD \cite{du2024dfadd}, and LibriSeVoc \cite{sun2023ai}. This partition allows us to systematically evaluate both in-domain performance and cross-domain generalization capability.

\subsection{Auxiliary Labels}

In this work, the design of auxiliary labels plays a crucial role in guiding the learning behavior of different components in our framework. For multitask learning, we construct auxiliary labels by combining dataset identity with spoof/bona fide annotations, forming joint labels of the form \textit{dataset $\times$ spoof/bona fide}. As a result, the model is encouraged to learn distinctions across datasets and spoofing conditions, thereby enhancing its discriminative capability in multi-domain scenarios. In contrast, for the GRL-based method, we employ only the dataset identity as the auxiliary label. The motivation is that GRL aims to learn domain-invariant representations. If spoof/bona fide information is also included in the auxiliary labels, the adversarial training may remove useful cues for spoof detection. Therefore, we use dataset-only labels so that GRL focuses on eliminating domain differences while preserving spoof-related information.

\subsection{Training Setup}

We adopt a two-stage training strategy to stabilize optimization and improve representation learning.

\textbf{Data processing.} For both stages, we apply data augmentation including reverberation and noise perturbation, with up to six augmented variants per sample. Training samples are generated using a class-weighted random segment sampler with a fixed chunk length of 2 seconds, ensuring balanced sampling across different datasets. For validation, we use a bucketing sampler to handle variable-length inputs efficiently.

\textbf{Stage 1.} In the first stage, we freeze the SSL feature extractor and train only the fuser, downstream backbone, and classifiers, which allows the model to learn stable representations without disrupting the pretrained features. During this stage, we use stochastic gradient descent (SGD) with a relatively large learning rate (0.4) and exponential decay scheduling. The effective batch size is set to 1024, and the model is trained with mixed-precision training. 

\textbf{Stage 2.} In the second stage, we unfreeze the entire model and perform full training. The learning rate is reduced to $5\times10^{-3}$ with a similar exponential decay schedule to ensure stable optimization. The effective batch size is set to 512, and we adopt the additive angular margin softmax (AAM-Softmax) loss with scale $s=32$ and margin $m=0.2$ for the main classification task.

In both stages, the auxiliary classifier or adapter is trainable, and we employ early stopping to prevent overfitting. Additionally, for both MT and GRL, we set the auxiliary loss weight $\lambda$ to 0.1 based on preliminary experiments on a smaller development subset.

% \begin{table}[t]
% \centering
% \caption{Evaluation results on the 2025 Speech Deepfake Arena benchmark. Results are reported in terms of EER (\%), where lower is better. The best result for each dataset is shown in bold.}
% \label{tab:arena_results_vertical}
% \small
% \setlength{\tabcolsep}{4pt}
% \renewcommand{\arraystretch}{1.08}
% \begin{tabularx}{\columnwidth}{@{}>{\raggedright\arraybackslash}Xccc@{}}
% \toprule
% \textbf{Dataset} & \textbf{Baseline} & \textbf{MT} & \textbf{GRL} \\
% \midrule
% In the Wild          & 5.211  & 4.778  & \textbf{3.062} \\
% ASV2019              & 5.850  & \textbf{1.526} & 4.453 \\
% ASV2021LA            & 10.940 & \textbf{6.990} & 13.710 \\
% ASV2021DF            & 5.261  & \textbf{4.041} & 5.042 \\
% ASV2024-EVAL         & \textbf{13.260} & 14.370 & 13.540 \\
% Fake or Real         & 1.428  & \textbf{0.057} & 4.583 \\
% Codecfake            & 19.990 & \textbf{15.080} & 18.980 \\
% ADD 2022 Track 1     & 23.950 & \textbf{21.910} & 22.570 \\
% ADD 2022 Track 3     & 5.274  & \textbf{4.535} & 5.718 \\
% ADD 2023 R1          & \textbf{15.980} & 19.970 & 18.390 \\
% ADD 2023 R2          & 23.920 & 21.230 & \textbf{18.290} \\
% DFADD                & 0.747  & \textbf{0.094} & 0.213 \\
% LibriSeVoc           & \textbf{0.056} & 0.115 & 0.661 \\
% SONAR                & 0.885  & \textbf{0.632} & 2.097 \\
% \midrule
% Average EER          & 9.482  & \textbf{8.219} & 9.379 \\
% Pooled EER           & 12.600 & 13.050 & \textbf{11.930} \\
% \bottomrule
% \end{tabularx}
% \end{table}
\begin{table}[t]
\centering
\caption{Evaluation results on the 2025 Speech Deepfake Arena benchmark. Results are reported in terms of EER (\%), where lower is better. The MT system uses Dataset Type $\times$ Spoof Detection labels as auxiliary supervision, while the GRL system uses Dataset Type labels for adversarial training. }
\label{tab:arena_results_vertical}
\small
\setlength{\tabcolsep}{4pt}
\renewcommand{\arraystretch}{1.08}
\begin{tabularx}{\columnwidth}{@{}>{\raggedright\arraybackslash}Xccc@{}}
\toprule
\textbf{Dataset} & \textbf{Baseline} & \textbf{MT} & \textbf{GRL} \\
\midrule
In the Wild              & 5.211  & 4.778  & \textbf{3.062} \\
ASV2019                  & 5.850  & \textbf{1.526} & 4.453 \\
ASV2021LA                & 10.940 & \textbf{6.986} & 13.712 \\
ASV2021DF                & 5.261  & \textbf{4.041} & 5.043 \\
ASV2024-EVAL             & \textbf{13.263} & 14.371 & 13.538 \\
Fake or Real             & 1.428  & \textbf{0.057} & 4.583 \\
Codecfake                & 19.990 & \textbf{15.084} & 18.980 \\
ADD 2022 Track 1         & 23.951 & \textbf{21.912} & 22.570 \\
ADD 2022 Track 3   & 5.275  & \textbf{4.535} & 5.718 \\
ADD 2023 R1              & \textbf{15.978} & 19.967 & 18.389 \\
ADD 2023 R2              & 23.935 & 21.233 & \textbf{18.290} \\
DFADD                    & 0.747  & \textbf{0.095} & 0.214 \\
LibriSeVoc               & \textbf{0.056} & 0.116 & 0.662 \\
SONAR                    & 0.885  & \textbf{0.632} & 2.097 \\
\midrule
Average EER              & 9.484  & \textbf{8.238} & 9.379 \\
Pooled EER               & 12.596 & 13.050 & \textbf{11.926} \\
\bottomrule
\end{tabularx}

\vspace{0.5mm}
\footnotesize
\textit{Note:} ASV2019 denotes ASVspoof 2019; ASV2021LA and ASV2021DF denote the LA and DF tracks of ASVspoof 2021, respectively; ASV2024-EVAL denotes the evaluation set of ASVspoof 2024.
\end{table}

% \begin{table}[t]
% \centering
% \caption{Comparison with top-performing systems on the 2025 Speech Deepfake Arena benchmark. Results are reported in terms of EER (\%), where lower is better.}
% \label{tab:arena_comparison}
% \scriptsize
% \setlength{\tabcolsep}{3pt}
% \renewcommand{\arraystretch}{1.08}
% \begin{tabularx}{\columnwidth}{@{}>{\raggedright\arraybackslash}Xccc@{}}
% \toprule
% \textbf{System} & \textbf{Params (M)} & \textbf{Pooled} & \textbf{Average} \\
% \midrule
% DF\_Arena\_1B\_V\_1       & 1000   & \textbf{9.524} & \textbf{5.919} \\
% DF\_Arena\_500M\_V\_1     & 500    & 10.880 & 5.780 \\
% XLSR+SLS                  & 340    & 16.079 & 14.015 \\
% TCM                       & 319    & 16.691 & 15.846 \\
% BiCrossMamba-ST           & 318.21 & 17.154 & 15.778 \\
% Nes2NetX                  & 317.9  & 17.366 & 16.186 \\
% Waav2Vec2\_AASIST         & 317.84 & 19.607 & 18.138 \\
% XLSR\_Mamba               & 319    & 20.591 & 14.647 \\
% Whisper\_Mesonet          & 7.6    & 23.551 & 28.355 \\
% \midrule
% \textbf{MT (Ours)}        & 315.4  & 13.050 & 8.238 \\
% \textbf{GRL (Ours)}       & 315.4  & 11.926 & 9.379 \\
% \bottomrule
% \end{tabularx}
% \end{table}

\begin{table}[t]
\centering
\caption{Comparison with top-performing open systems on the 2025
Speech Deepfake Arena benchmark. Params denotes the number of parameters
in millions. Results are reported as EER (\%), where lower is better.
The best, second-best, and third-best results are indicated by bold,
italics, and underline, respectively.}
\label{tab:arena_comparison}

\footnotesize
\setlength{\tabcolsep}{2.5pt}
\renewcommand{\arraystretch}{1.08}

\begin{tabularx}{\columnwidth}{@{}>{\raggedright\arraybackslash}Xrrr@{}}
\toprule
\textbf{System} & \textbf{Params (M)} & \textbf{Avg.} & \textbf{Pooled} \\
\midrule
DF\_Arena\_1B\_V\_1~\cite{kulkarni2026compactsslbackbonesmatter}
& 1000 & \textit{5.919} & \textbf{9.524} \\

DF\_Arena\_500M\_V\_1~\cite{kulkarni2026compactsslbackbonesmatter}
& 500 & \textbf{5.780} & \textit{10.880} \\

XLSR+SLS~\cite{zhang2024audio}
& 340 & 14.015 & 16.079 \\

TCM~\cite{truong2024temporal}
& 319 & 15.846 & 16.691 \\

BiCrossMamba-ST~\cite{kheir2025bicrossmamba}
& 318.2 & 15.778 & 17.154 \\

Nes2NetX~\cite{liu2025nes2net}
& 317.9 & 16.186 & 17.366 \\

Wav2Vec2\_AASIST~\cite{tak2022automatic}
& 317.8 & 18.138 & 19.607 \\

XLSR\_Mamba~\cite{xiao2025xlsr}
& 319 & 14.647 & 20.591 \\

Whisper\_Mesonet~\cite{kawa2023improved}
& 7.6 & 28.355 & 23.551 \\
\midrule
\textbf{MT (Ours)}
& 315.4 & \underline{8.238} & 13.050 \\

\textbf{GRL (Ours)}
& 315.4 & 9.379 & \underline{11.926} \\
\bottomrule
\end{tabularx}
\end{table}

\begin{table}[t]
\centering
\caption{Ablation study on auxiliary label design. Results are reported in terms of EER (\%), where lower is better. The best result within each method group is shown in bold.}
\label{tab:ablation_aux_label}
\small
\setlength{\tabcolsep}{3pt}
\renewcommand{\arraystretch}{1.08}
\begin{tabularx}{\columnwidth}{@{}l
>{\centering\arraybackslash}X
>{\centering\arraybackslash}X
!{\vrule width 0.4pt}
>{\centering\arraybackslash}X
>{\centering\arraybackslash}X@{}}
\toprule
\textbf{Dataset} 
& \multicolumn{2}{c}{\textbf{MT}} 
& \multicolumn{2}{c}{\textbf{GRL}} \\
\cmidrule(lr){2-3} \cmidrule(lr){4-5}
& \textbf{D} 
& \textbf{D$\times$S}
& \textbf{Lang.}
& \textbf{D} \\
\midrule
In the Wild      & \textbf{3.072} & 4.778  & 3.276  & \textbf{3.062} \\
ASV2019          & 7.518  & \textbf{1.526}  & 9.353  & \textbf{4.453} \\
ASV2021LA        & 16.037 & \textbf{6.986}  & 15.697 & \textbf{13.712} \\
ASV2021DF        & 5.706  & \textbf{4.041}  & \textbf{4.644}  & 5.043 \\
ASV24-EVAL       & \textbf{13.506} & 14.371 & \textbf{11.512} & 13.538 \\
Fake or Real     & 1.165  & \textbf{0.057}  & \textbf{2.289}  & 4.583 \\
Codecfake        & 17.655 & \textbf{15.084} & 24.398 & \textbf{18.980} \\
ADD22-T1         & 28.283 & \textbf{21.912} & 24.347 & \textbf{22.570} \\
ADD22-T3         & 5.312  & \textbf{4.535}  & 6.176  & \textbf{5.718} \\
ADD 2023 R1      & \textbf{16.101} & 19.967 & 20.745 & \textbf{18.389} \\
ADD 2023 R2      & 26.695 & \textbf{21.233} & 29.856 & \textbf{18.290} \\
DFADD            & \textbf{0.000} & 0.095  & 1.305  & \textbf{0.214} \\
LibriSeVoc       & \textbf{0.038} & 0.116  & \textbf{0.049} & 0.662 \\
SONAR            & 0.862  & \textbf{0.632}  & 3.034  & \textbf{2.097} \\
\midrule
Average EER      & 10.139 & \textbf{8.238}  & 11.192 & \textbf{9.379} \\
Pooled EER       & 14.762 & \textbf{13.050} & 12.939 & \textbf{11.926} \\
\bottomrule
\end{tabularx}

\vspace{0.5mm}
\footnotesize
\textit{Note:} D denotes Dataset, S denotes Spoof\_det, and Lang. denotes Language. ASV2019: ASVspoof 2019; ASV2021LA \& ASV2021DF: the LA and DF tracks of ASVspoof 2021, respectively; 
ASV24-EVAL: the evaluation set of ASVspoof 2024; ADD22-T1: ADD 2022 Track 1; ADD22-T3: ADD 2022 Track 3.
\end{table}

\begin{figure*}[!t]
\centering
\captionsetup[subfigure]{skip=1pt}
\captionsetup{skip=3pt}

% ```
% \makebox[\textwidth][c]{%
%     \makebox[0.43\textwidth][c]{\textbf{Dataset\_type}}%
%     \makebox[0.43\textwidth][c]{\textbf{Spoof\_det}}%
% }

% \vspace{0.05em}

\begin{subfigure}{\textwidth}
    \centering
    \includegraphics[width=0.41\linewidth]{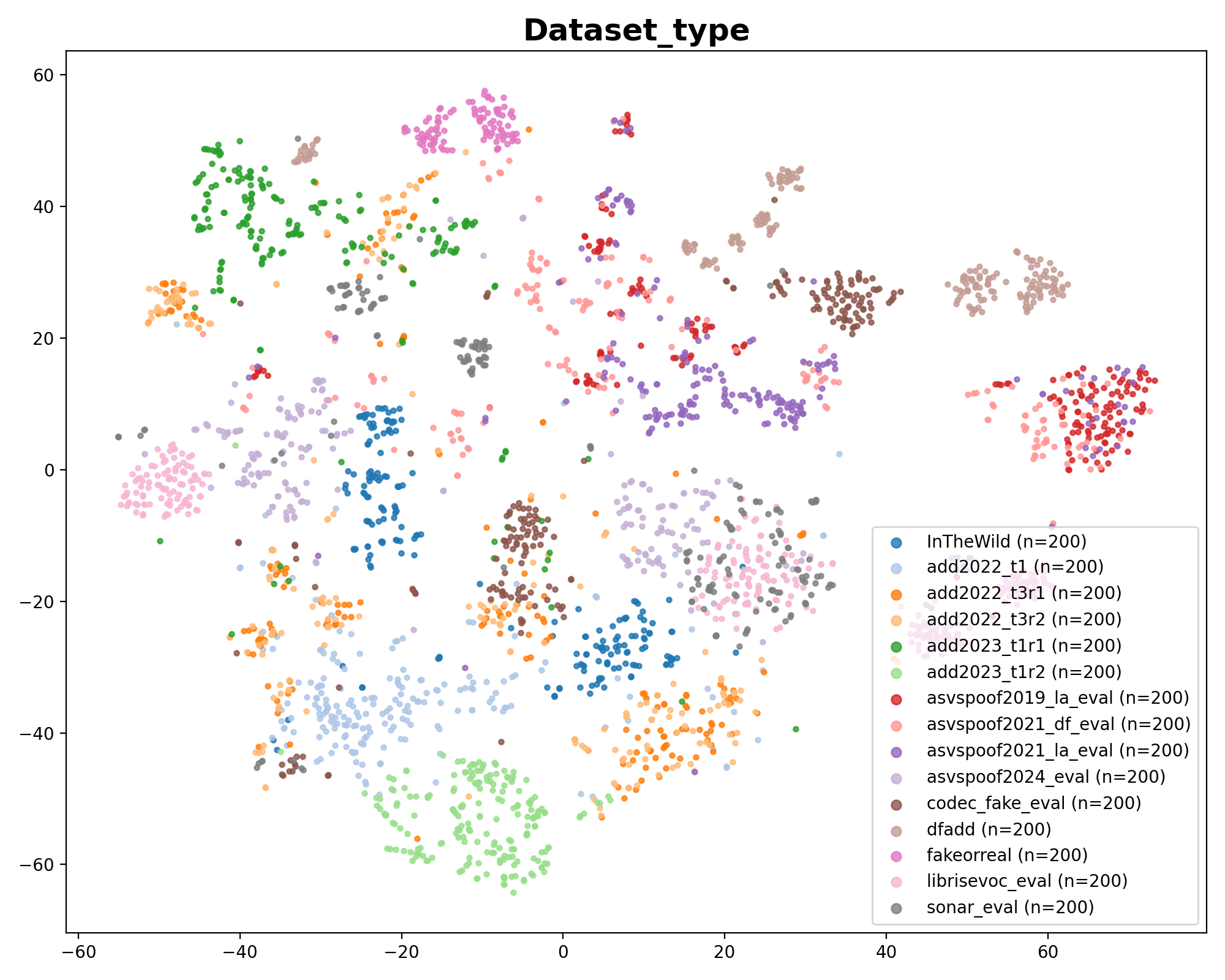}%
    \hspace{0.025\linewidth}
    \includegraphics[width=0.41\linewidth]{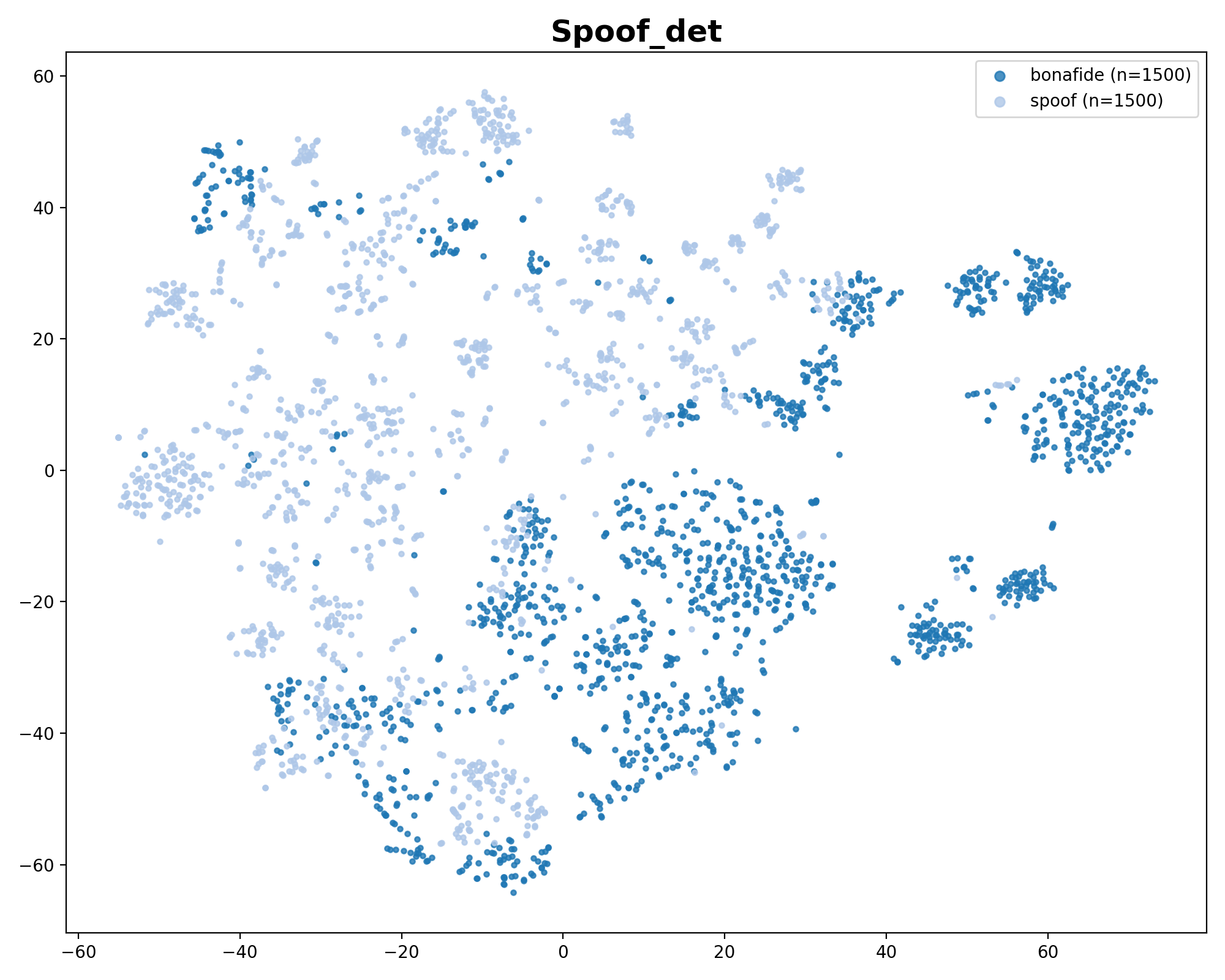}
    \caption{Baseline}
    \label{fig:tsne_baseline}
\end{subfigure}

\vspace{0.05em}

\begin{subfigure}{\textwidth}
    \centering
    \includegraphics[width=0.41\linewidth]{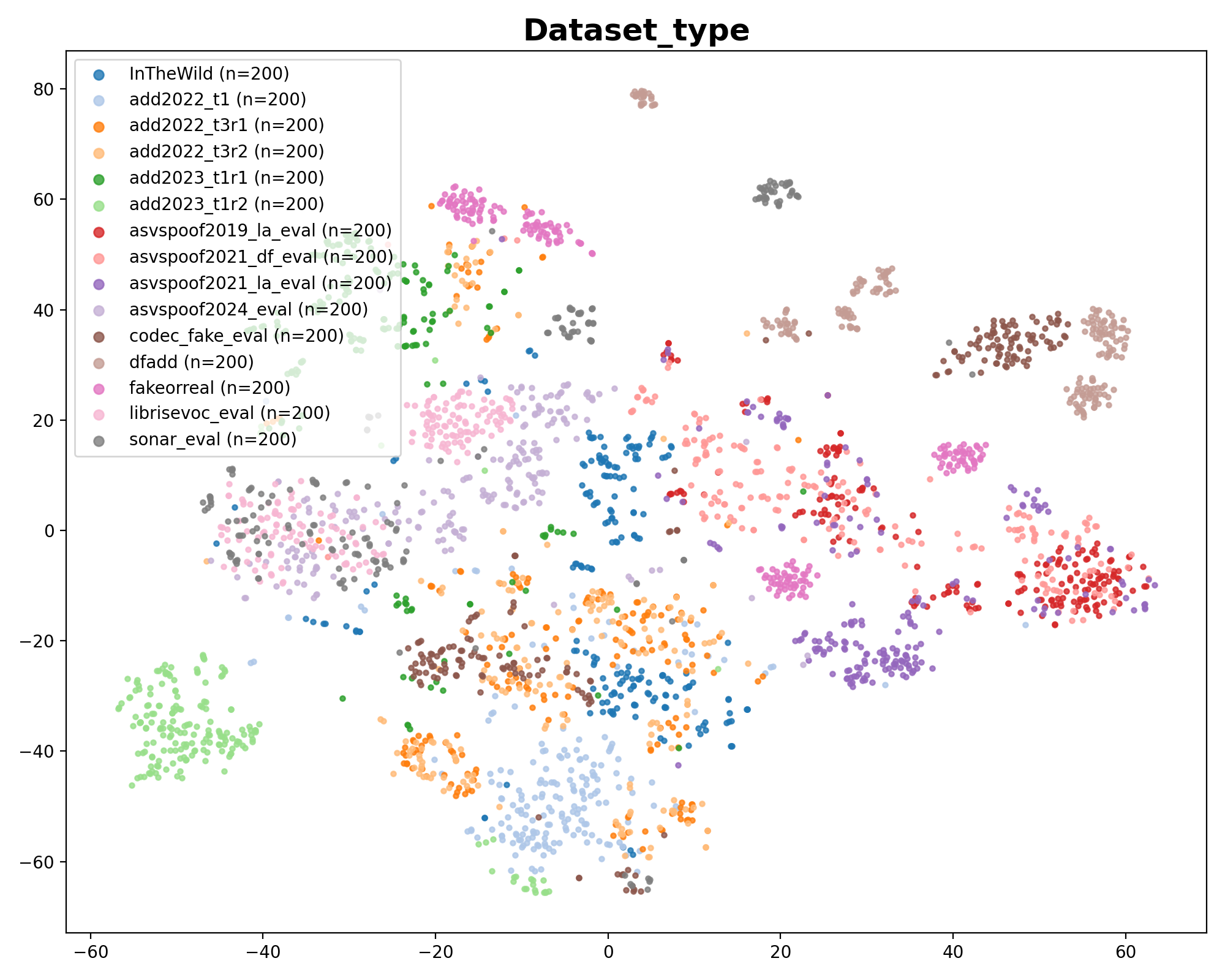}%
    \hspace{0.025\linewidth}
    \includegraphics[width=0.41\linewidth]{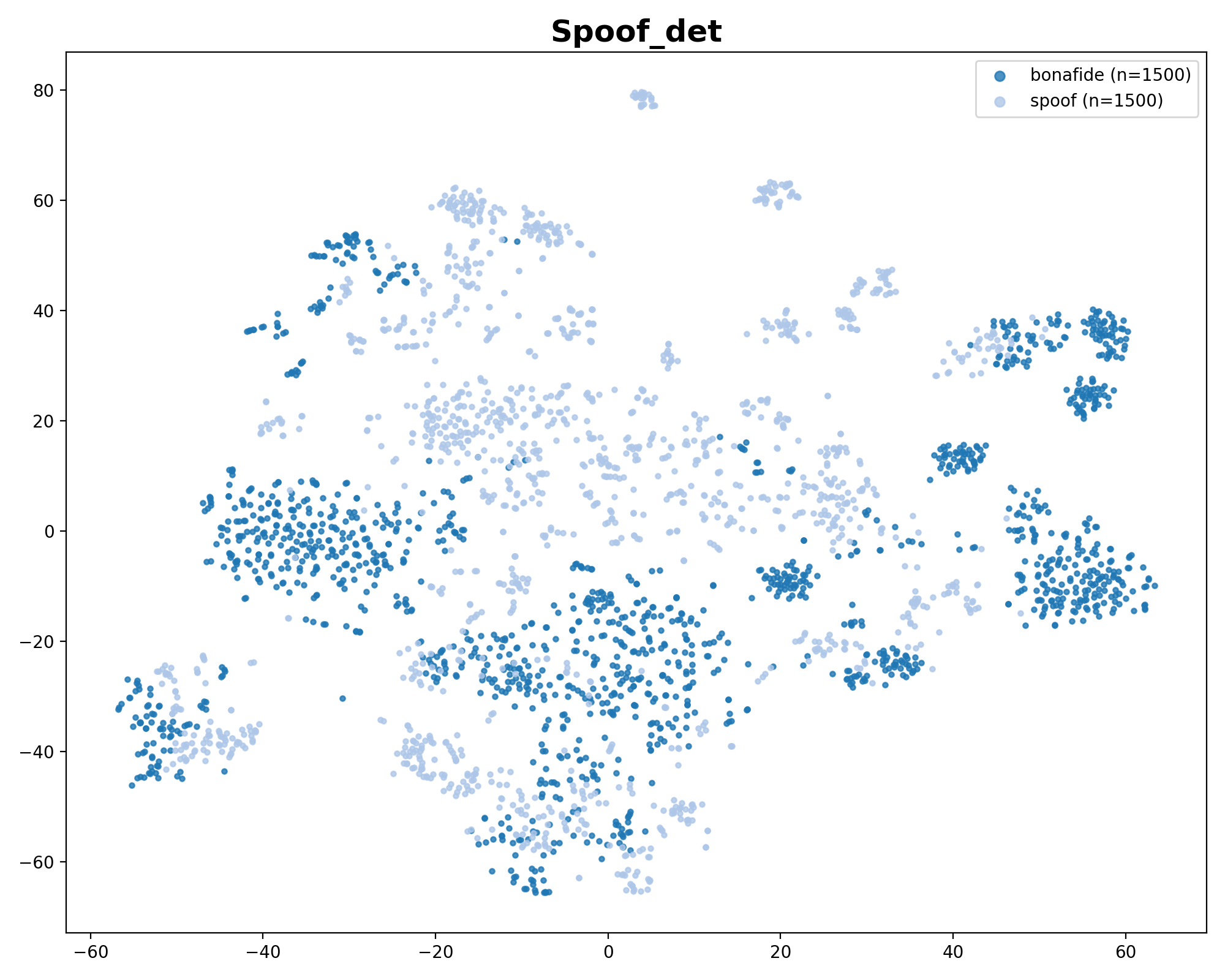}
    \caption{MT}
    \label{fig:tsne_mt}
\end{subfigure}

\vspace{0.05em}

\begin{subfigure}{\textwidth}
    \centering
    \includegraphics[width=0.41\linewidth]{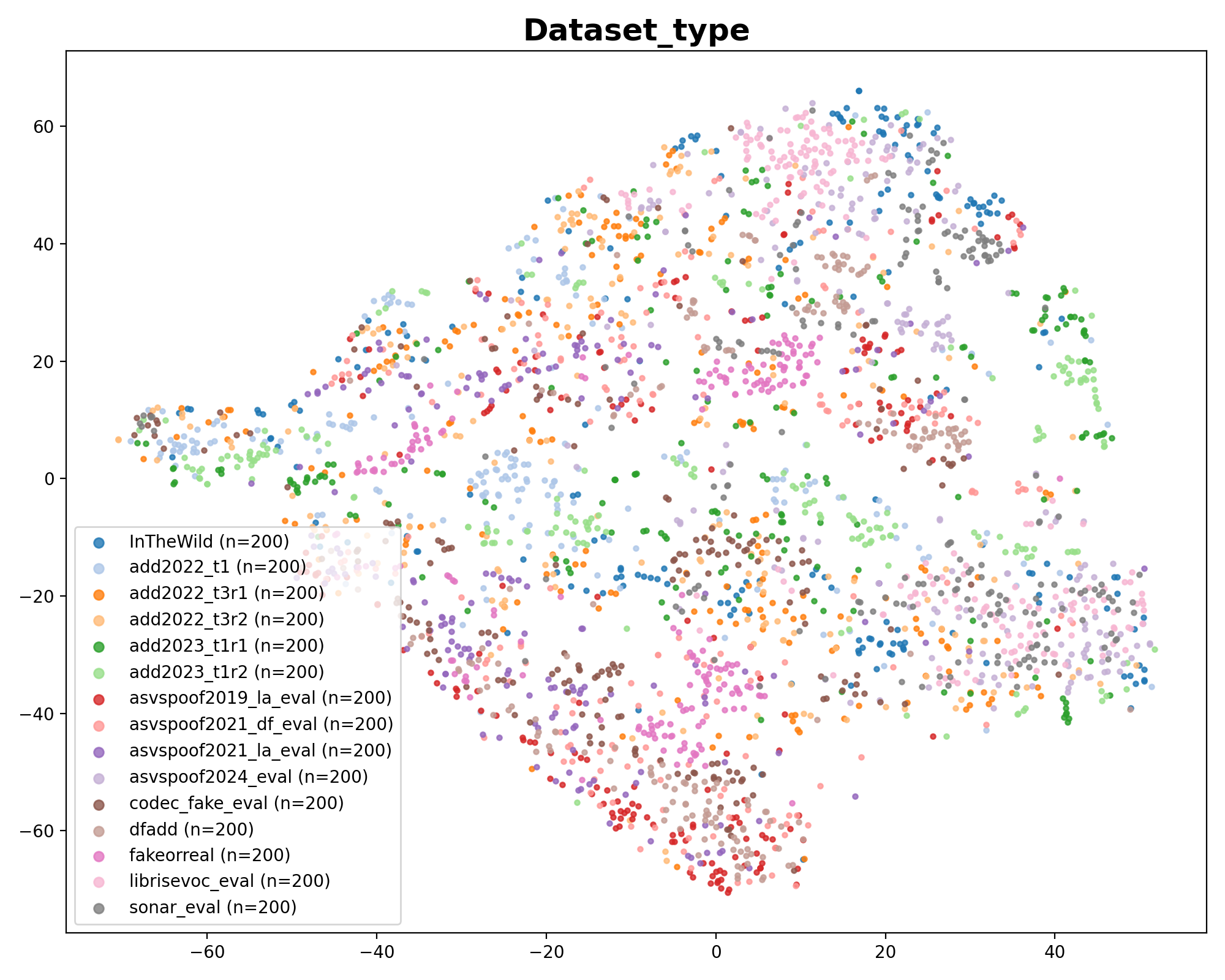}%
    \hspace{0.025\linewidth}
    \includegraphics[width=0.41\linewidth]{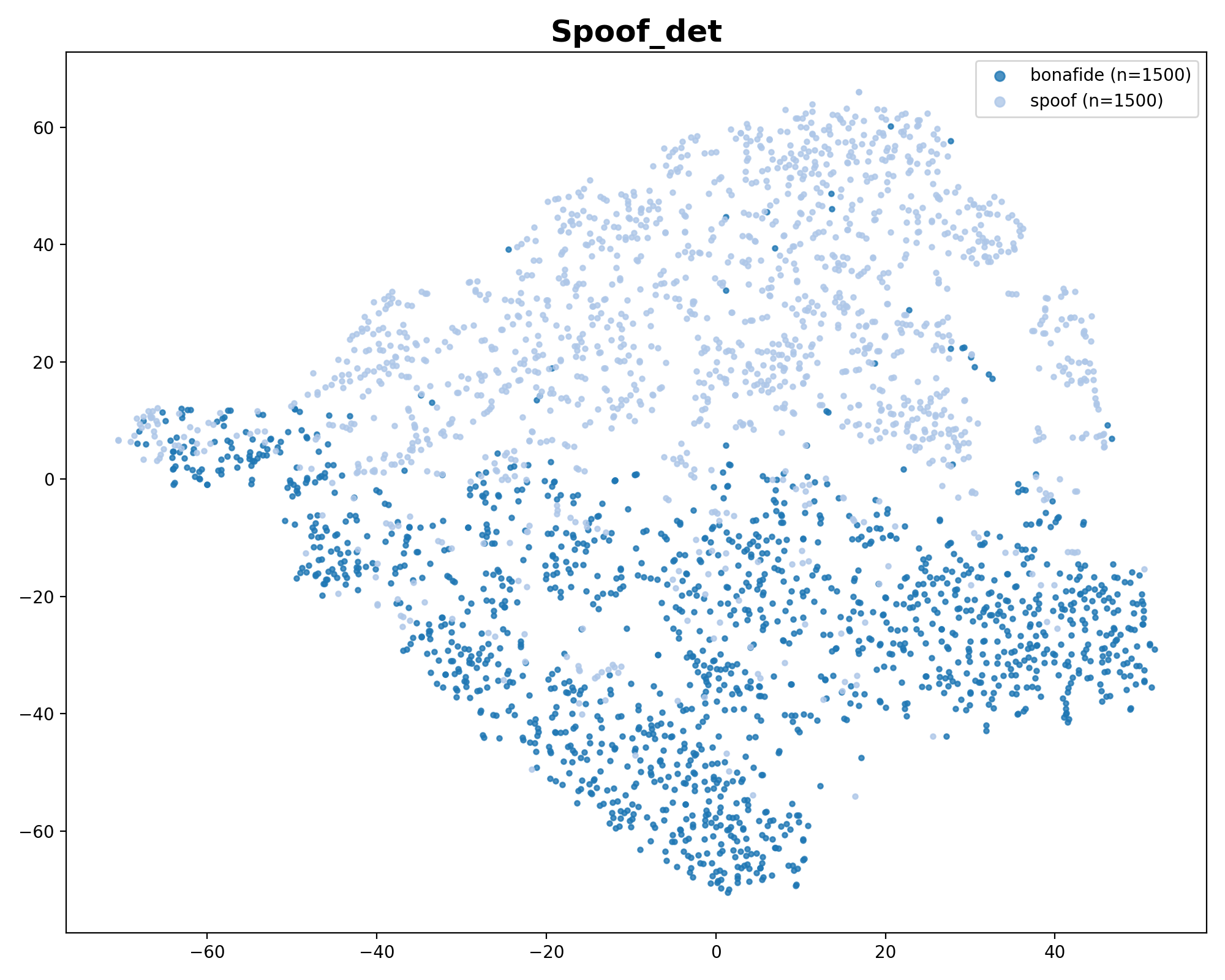}
    \caption{GRL}
    \label{fig:tsne_grl}
\end{subfigure}

\vspace{-0.3em}
\caption{t-SNE visualization of learned embeddings under different training strategies. Each row corresponds to one model configuration: Baseline, MT, and GRL. Within each row, the two panels show the same embedding space colored by Dataset\_type labels and Spoof\_det labels, respectively.}
\label{fig:tsne_visualization}
```

\end{figure*}

\section{Results and Analysis}

% \vspace{0.5em}

\subsection{MT \& GRL Results}
\label{sec:grl_mt_results}

Table~\ref{tab:arena_results_vertical} presents the evaluation results of the baseline, multitask learning (MT), and gradient reversal layer (GRL) methods on the 2025 Speech Deepfake Arena benchmark. Overall, MT achieves the best Average EER, reducing the baseline from 9.484\% to 8.238\%, corresponding to a relative improvement of 13.14\%. This indicates that explicitly introducing auxiliary domain-related supervision improves the average detection performance across diverse evaluation datasets.
% This indicates that explicitly introducing auxiliary domain-related supervision helps the model learn more discriminative representations across diverse evaluation datasets.
% At the time of submission, this result ranks 10th on the Speech Deepfake Arena leaderboard\footnote{\url{https://huggingface.co/spaces/Speech-Arena-2025/Speech-DF-Arena}} in terms of Average EER, and 3rd among open-source systems.

At the dataset level, MT obtains the best performance on 9 out of 14 evaluation subsets, including ASVspoof 2019, ASVspoof 2021LA, ASVspoof 2021DF, Fake or Real, Codecfake, ADD 2022 Track 1, ADD 2022 Track 3, DFADD, and SONAR. In particular, MT leads to substantial improvements on several challenging subsets, such as ASVspoof 2019, where the EER decreases from 5.850\% to 1.526\%, and Fake or Real, where the EER decreases from 1.428\% to 0.057\%. These results suggest that the MT objective can effectively exploit dataset-level supervision to improve generalization across diverse evaluation datasets.
% These results suggest that the MT objective can effectively exploit dataset-level supervision to improve cross-dataset robustness.

Compared with MT, GRL shows a different behavior. Although GRL does not achieve the best Average EER, it obtains the best Pooled EER, reducing the baseline from 12.596\% to 11.926\%, corresponding to a relative improvement of 5.32\%. GRL also achieves the best result on In the Wild and ADD 2023 R2. This suggests that GRL may be more beneficial when aggregating trials across datasets under a shared decision threshold, potentially due to improved cross-dataset score consistency.
% This suggests that adversarial domain-invariant learning may be more beneficial when aggregating trials across datasets, especially under distribution mismatch.
% At the time of submission, this result ranks 11th on the Speech Deepfake Arena leaderboard in terms of Pooled EER, and 3rd among open-source systems.

Table~\ref{tab:arena_comparison} further compares our systems with top-performing open systems on the 2025 Speech Deepfake Arena leaderboard. Among the open systems summarized in the table, MT achieves the third-best Average EER, while GRL achieves the third-best Pooled EER. Notably, the two systems that outperform ours use substantially larger models, with 500M and 1000M parameters, whereas our MT and GRL systems use 315.4M parameters. This comparison suggests that the proposed training strategies provide a favorable trade-off between detection performance and model size, achieving competitive leaderboard performance without relying on substantially larger model capacity.

These observations reveal distinct performance profiles for MT and GRL. MT improves the average performance across individual datasets through explicit, dataset-aware auxiliary supervision, whereas GRL yields better pooled performance under an objective designed to reduce dataset-specific information. Therefore, MT is more effective in improving the macro-average of per-dataset EERs, whereas GRL can be advantageous for pooled evaluation across heterogeneous datasets.
% Therefore, MT is more effective in improving per-dataset generalization, whereas GRL can be advantageous for global robustness under heterogeneous evaluation conditions.

\subsection{Visualization Analysis}
\label{sec:vis_analysis}

Fig.~\ref{fig:tsne_visualization} shows the t-SNE visualization of the learned embeddings extracted after the backbone pooling layer. For each training strategy, the two panels in the same row correspond to the same embedding space, but are colored according to Dataset\_type, which denotes the source dataset category, and Spoof\_det labels, which denotes the binary bona fide/spoof label, respectively. The visualized samples are drawn from the evaluation split, with an equal number of bona fide and spoof samples sampled from each dataset. This visualization provides qualitative evidence for how different training objectives affect the structure of the learned representation.

For the baseline model, the embeddings exhibit noticeable dataset-dependent clustering. Samples from the same dataset tend to occupy similar local regions, indicating that the baseline representation contains strong dataset-specific information. However, when colored by Spoof\_det labels, bona fide and spoof samples are not clearly separated into two global clusters. This suggests that the baseline model may partially rely on dataset-related biases rather than learning purely spoof-discriminative cues.

Compared with the baseline, MT produces a more structured feature space with respect to Dataset\_type labels. Samples from the same dataset form clearer local clusters, which is consistent with the design of the MT objective. Since MT explicitly introduces an auxiliary task to model dataset-related information, the backbone is encouraged to preserve domain-dependent variations. Meanwhile, the Spoof\_det visualization shows that MT does not simply force all bona fide and spoof samples into two global clusters. Instead, it appears to learn more fine-grained spoof-discriminative structures within different dataset regions. This observation is consistent with the quantitative results in Table~\ref{tab:arena_results_vertical}, where MT achieves the best Average EER and performs best on most individual evaluation subsets.

In contrast, GRL leads to a substantially more mixed embedding space across datasets. In the Dataset\_type visualization, samples from different datasets are more entangled, and the dataset-specific clusters observed in the baseline and MT settings become less pronounced. This indicates that the adversarial objective imposed by GRL suppresses dataset-discriminative information from the backbone representation. As a result, GRL encourages a more domain-invariant feature space, where samples from heterogeneous datasets are better aligned. In the Spoof\_det visualization, GRL also shows a more globally organized bona fide/spoof structure, suggesting that reducing dataset separability can help the model focus on spoof-related information shared across datasets. This is consistent with the Pooled EER results in Table~\ref{tab:arena_results_vertical}, where GRL achieves the best pooled performance.

Overall, the visualization reveals a clear difference between the two strategies. MT learns domain-aware representations by explicitly modeling dataset-specific variations, which benefits per-dataset performance. GRL, on the other hand, learns domain-invariant representations by reducing dataset separability, which improves global generalization under heterogeneous evaluation conditions. These qualitative findings support the complementary behavior observed in the quantitative results.

\subsection{Ablation Study Result}
\label{sec:ablation_study}

Table~\ref{tab:ablation_aux_label} presents the ablation study on different auxiliary label designs for MT and GRL. For MT, we compare two auxiliary supervision strategies: dataset-only labels and Dataset\_type $\times$ Spoof\_det labels. For GRL, we compare language-based adversarial labels with dataset-based adversarial labels.

While prior MT-based studies typically use a single auxiliary label, such as language or codec type, our Dataset\_type $\times$ Spoof\_det design jointly encodes dataset identity and bona fide/spoof information. We therefore compare it with a dataset-only label design to verify the benefit of this combined auxiliary supervision. As shown in Table~\ref{tab:ablation_aux_label}, using Dataset\_type $\times$ Spoof\_det labels consistently outperforms using dataset-only labels on most evaluation subsets. Specifically, Dataset\_type $\times$ Spoof\_det achieves better results on 9 out of 14 datasets and reduces the Average EER from 10.139\% to 8.238\%. The improvement is especially clear on ASVspoof 2019, Fake or Real, and ADD 2023 R2, where incorporating spoof/bona fide information into the auxiliary label provides substantially stronger supervision. These results indicate that simply modeling dataset identity is not sufficient for MT. Instead, combining dataset identity with spoof\_det labels allows the auxiliary task to capture both domain variation and spoof-related distribution differences, leading to more discriminative representations.

For the GRL setting, we further investigate the choice of adversarial labels. Previous studies have also explored using language or codec-related attributes as auxiliary labels for adversarial training. However, these settings are usually conducted on one or two datasets where such auxiliary annotations are naturally available. In contrast, our training setup mixes multiple datasets collected from different sources, making it difficult to obtain consistent language or codec labels for all samples. In particular, codec-type annotations are not reliably available in our collected training data. Therefore, we focus on comparing dataset-based adversarial labels with language-based adversarial labels. To construct the language labels, we use a pre-trained VoxLingua107 ECAPA-TDNN spoken language identification model\footnote{\url{https://huggingface.co/speechbrain/lang-id-voxlingua107-ecapa}} from SpeechBrain \cite{speechbrain} to predict the language of each training sample. This model achieves an error rate of 6.7\% on the VoxLingua107 development set, providing reasonably reliable language predictions for constructing pseudo language labels at scale. The predicted language labels are then used as pseudo labels for the GRL auxiliary classifier.

As shown in Table~\ref{tab:ablation_aux_label}, dataset-based adversarial training achieves better aggregate performance than language-based adversarial training. GRL with dataset labels also performs better across most individual datasets. Compared with language labels, it reduces the Average EER from 11.192\% to 9.379\% and the Pooled EER from 12.939\% to 11.926\%, corresponding to relative reductions of 16.20\% and 7.83\%, respectively. This suggests that dataset identity provides a more effective adversarial signal than language identity for encouraging domain-invariant feature learning in this benchmark. Since the evaluation datasets differ not only in language but also in recording conditions, spoofing algorithms, data collection protocols, and corpus-specific artifacts, dataset labels can capture broader domain discrepancies than language labels alone.

% Overall, the ablation results show that the choice of auxiliary label is critical for both MT and GRL. For MT, Dataset $\times$ Spoof labels are more effective because they explicitly model domain-dependent spoof characteristics. For GRL, dataset labels provide a stronger adversarial objective than language labels, leading to better overall robustness, especially under pooled evaluation. These findings support the use of Dataset $\times$ Spoof labels for MT and dataset labels for GRL in our final system.

\section{Conclusion}
\label{sec:conclusion}

In this work, we propose a dataset-aware framework to improve audio deepfake detection across heterogeneous evaluation settings. Instead of relying on auxiliary annotations such as language, codec, or spoofing method, which are often unavailable or inconsistent across datasets, our method uses dataset identity as a naturally available supervisory signal. Based on this idea, we explore two distinct strategies: multitask learning with Dataset\_type $\times$ Spoof\_det labels and adversarial training with dataset-based GRL.

% Experimental results on the 2025 Speech Deepfake Arena benchmark demonstrate that the proposed framework improves cross-dataset robustness. 
Experimental results on the 2025 Speech Deepfake Arena benchmark show distinct performance gains for MT and GRL under Average EER and Pooled EER, respectively. MT achieves the best Average EER, reducing it from 9.484\% to 8.238\%, while GRL achieves the best Pooled EER, reducing it from 12.596\% to 11.926\%. The t-SNE visualization qualitatively suggests that MT encourages domain-aware representations, whereas GRL promotes more domain-invariant feature alignment. In addition, the ablation study confirms that Dataset\_type $\times$ Spoof\_det labels are more effective than dataset-only labels for MT, and dataset labels provide a stronger adversarial signal than language pseudo labels for GRL.

Overall, these results show that dataset identity can serve as a simple and effective supervisory signal for improving aggregate detection performance across heterogeneous evaluation datasets.
% Overall, these results show that dataset identity can serve as a simple, scalable, and effective signal for improving audio deepfake detection under mixed-dataset and in-the-wild settings. 
% Future work will explore how to further combine domain-aware and domain-invariant learning strategies within a unified training objective.

\section{Limitation \& Future Work}
\label{sec:limitation_future}

Despite the promising results, this work still has several limitations. First, MT and GRL are investigated as two separate strategies. As shown in the results, MT achieves better Average EER by learning domain-aware representations, while GRL achieves better Pooled EER by encouraging domain-invariant representations. However, we have not yet developed a unified framework that can effectively combine these two complementary behaviors.

Second, our experiments are conducted based on one backbone architecture. Although the proposed auxiliary label designs improve performance in this setting, it remains unclear whether the same MT and GRL modules can consistently benefit other deepfake detection backbones, such as different SSL encoders, pooling strategies, or classifier architectures.

For future work, we plan to explore a unified training objective that jointly leverages the strengths of MT and GRL, allowing the model to adaptively balance domain-aware and domain-invariant learning. In addition, we will evaluate the proposed dataset-aware training strategies on more backbone architectures to verify their general applicability and robustness across different model designs.

\section{Acknowledgments}

This work was supported in part by computational resources provided by the Center for Language and Speech Processing (CLSP) at Johns Hopkins University. We thank Jan Vainer from Meaning for his helpful assistance.

% {\color{blue}Acknowledgments should be included only in the camera-ready version, not in the version submitted for review. For regular papers, pages 5 and 6, and for long papers, pages 9 and 10, are reserved exclusively for acknowledgments, disclosures of the use of generative AI tools, and references. No other content may appear on these pages. Any appendices must be contained within the first four pages for regular papers and within the first eight pages for long papers.

% Acknowledgments and references may begin on an earlier page if space permits.}

% \ifcameraready
%      The Interspeech 2026 organizers
% \else
%      The authors
% \fi
% would like to thank ISCA and the organizing committees of past Interspeech conferences for their help and for kindly providing the previous version of this template.

\section{Generative AI Use Disclosure}
The author used generative AI tools to assist with language polishing, grammar correction, writing organization, and LaTeX formatting. All research ideas, experimental design, implementation, data analysis, results, and conclusions were developed and finalized by the author. The author takes full responsibility for the content of this work.

\bibliographystyle{IEEEtran}
\bibliography{mybib}

\end{document}